# Boron phosphide under pressure: *in situ* study by Raman scattering and X-ray diffraction

Vladimir L. Solozhenko,[1] Oleksandr O. Kurakevych,[2] Yann Le Godec,[2] Aleksandr V. Kurnosov,[3] and Artem R. Oganov[4]

[1] LSPM–CNRS, Université Paris Nord, 93430 Villetaneuse, France

[2] IMPMC, UPMC Sorbonne Universités, UMR CNRS 7590, Muséum National d'Histoire Naturelle, IRD UMR 206, 75005 Paris, France

[3] Bayerisches Geoinstitut, Universität Bayreuth, 95440 Bayreuth, Germany

[4] Department of Geosciences, Center for Materials by Design, Institute for Advanced Computational Science, Stony Brook University, Stony Brook, NY 11794-2100, USA

## Abstract

Cubic boron phosphide BP has been studied *in situ* by X-ray diffraction and Raman scattering up to 55 GPa at 300 K in a diamond anvil cell. The bulk modulus of $B_0 = 174(2)$ GPa has been established, which is in excellent agreement with our *ab initio* calculations. The data on Raman shift as a function of pressure, combined with equation-of-state data, allowed us to estimate the Grüneisen parameters of the TO and LO modes of zinc-blende structure, $\gamma_G^{TO} = 1.16$ and $\gamma_G^{LO} = 1.04$, just like in the case of other $A^{III}B^V$ diamond-like phases, for which $\gamma_G^{TO} > \gamma_G^{LO} \cong 1$. We also established that the pressure dependence of the effective electro-optical constant $\alpha$ is responsible for a strong change in relative intensities of the TO and LO modes from $I_{TO}/I_{LO} \sim 0.25$ at 0.1 MPa to $I_{TO}/I_{LO} \sim 2.5$ at 45 GPa, for which we also find excellent agreement between experiment and theory.

## I. INTRODUCTION

In recent years, interest in high-pressure behavior of solids composed of elements with low atomic numbers has sharply increased.[1] Novel compounds,[2,3] nanostructures,[4] exotic crystal structures[5] and unexpected solid solutions[6] have been discovered, with a potential for their use as advanced materials. Boron phosphides[7,8] still remain ill-understood[9-11] and very different from heavier analogs in terms of the bandgap,[11,12] electronic structure evolution and, especially, high-pressure stability.[13] In order to understand all these peculiarities, one needs reliable experimental data on compressibility and high-pressure lattice dynamics.

The combination of X-ray diffraction and Raman spectroscopy provides valuable insight into the lattice behavior of solids under pressure and allows understanding the thermodynamic, electronic, and optical properties of diamond-like structures.[14,15] In the present paper we report the results of Raman and X-ray diffraction studies of BP under pressures to 55 GPa at room temperature. The measured equation of state has been compared with *ab initio* calculations. The analysis of TO and LO Raman active modes allowed us to establish the Grüneisen parameters. The observed pronounced increase of relative TO-to-LO intensities have been explained by pressure dependence of the effective electro-optical constant, $\alpha$, using the model described in Ref.[15].

## II. EXPERIMENTAL

The BP powder sample was synthesized by magnesium thermal reduction method.[16] The amorphous $BPO_4$ precursor obtained by the reaction between boric ($H_3BO_3$) and phosphoric ($H_3PO_3$) acids has been reduced by metallic Mg in the mode of self-propagating high-temperature synthesis in the presence of sodium chloride NaCl (to avoid the overheating of a reaction mixture). The product was boiled for an hour in excess of 5N-hydrochloric acid, and then many times washed with distilled water, and dried in air at 50°C. The lattice parameter $a = 4.5356(9)$ Å of the as-synthesized single-phase BP agrees well with the literature data, e.g. $a = 4.538(2)$ Å.[17]

Single crystals of BP up to 100 μm were obtained using flux growth technique by interaction of phosphorus vapor with boron dissolved in nickel in a sealed (previously evacuated) quartz tube.[18] The boron-nickel solution was located at one end of the tube and held at 1150°C; with phosphorus at the opposite end at 430°C (to produce a vapor pressure to 0.5 MPa). Transparent red crystals were grown with a cooling rate of 3 K per hour. The lattice constant of the crystals was 4.534 Å, close to the literature data ($a = 4.538(2)$ Å[17]).

*In situ* X-ray diffraction and Raman experiments up to 55 GPa were conducted in the diamond anvil cells (DAC). We used type Ia diamonds in four-pin type DACs. The sample was loaded in a Ne pressure medium in a 150-mm hole of a preindented stainless steel or Re gasket and a tiny (~3 μm in diameter) ruby sphere chip was positioned near the center of the hole. The sample pressure was determined from both the shift of the ruby fluorescence $R_1$ line[19] and equations of state of gold[20] or neon.[21] The pressure-induced frequency shifts were monitored on compression to about 55 GPa. The absence of large broadening of ruby fluorescence lines, BP, neon and gold reflections in the X-ray diffraction patterns and BP Raman bands, reproducibility of the EOS data and Raman shift with pressure in different runs have indicated the relatively low strains and stresses in the samples over the whole pressure range studied here, as well as insignificant pressure gradients throughout the cell.

The Raman scattering measurements were performed at 300 K using a Dilor XY system with the 514.5 nm Ar$^+$ ion laser as the excitation source. The scattered light was collected in the backscattering geometry using a CCD detector. The incident laser power was 50–75 mW for the measurements in air, while for samples inside the DAC the power varied in the 250–300 mW range. The observed Raman bands did not change within the laser power range used in our experiments. The spectrometer was calibrated using the $\Gamma_{25}$ phonon of diamond Si (*Fd-3m*). The Pearson function was used to fit the peak profiles. At ambient conditions we estimate a resolution of 1 cm$^{-1}$ for the Raman peak position and 0.5 cm$^{-1}$ for the peak full width at half maximum.

The equation of state of BP was measured in a DAC up to 45 GPa using the in-house X-ray 3-circle Bruker diffractometer equipped with a SMART APEX CCD detector and a high-brilliance Rigaku rotating anode (Rotor Flex FR-D, Mo-Kα radiation) with Osmic focusing X-ray optics; and up to 55 GPa using synchrotron radiation and on-line large-area Bruker CCD detector at ID27 high-pressure beamline, ESRF. Correction of the two-dimensional diffraction images for spatial distortions and integration of the Debye–Scherrer rings were performed using the FIT2D software.

### III. COMPUTATIONAL METHODOLOGY

We have also theoretically computed the equation of state of BP. These calculations are based on density functional theory within the generalized gradient approximation (GGA),[22] using the all-electron projector-augmented wave (PAW) method[23,24] as implemented in the VASP code.[25] The PAW potentials that we used had [He] core (radius 1.7 a.u.) for boron atoms and [Ne] core (radius 1.9 a.u.) for phosphorus atoms. In these calculations we used plane-wave basis sets with the kinetic energy cut-off of 450 eV, and the Brillouin zone was sampled by an 8x8x8 Monkhorst-Pack mesh. The wavefunctions were optimized until the energies converged to better than 10$^{-5}$ eV/cell and the structure was relaxed at pressures from 0 to 100 GPa with pressure interval of 10 GPa. At each pressure relaxation proceeded until the enthalpy changes were below 10$^{-4}$ eV/cell.

The optimized unit cell parameter at 0.1 MPa is 4.546 Å , in remarkable agreement with the experimental value: usually theoretical and experimental cell parameters differ by ~1 %, but BP in this work, as well as pure boron[5] show much higher accuracy (~0.2 %). Part of this remarkable agreement is due to cancellation of errors, as the calculations were done at zero Kelvin, ignoring lattice vibrations and their effect on the cell parameter, but taking this effect into account would not worsen the agreement much. We should also mention here that at the correct volume such theoretical calculations give remarkably accurate predictions of the physical properties such as the elastic constants.[26] Since the unit cell volume is predicted accurately here, we can expect an accurate prediction of the equation of state.

## IV. RESULTS AND DISCUSSION

### A. Equation of state of BP.

At ambient conditions BP has a cubic zinc-blend structure (space group $F$-43$m$ (No. 216)) with unit cell parameter $a = 4.538(2)$ Å.[8,17] The aperture of employed diamond anvil cell have allowed us to observe lines *111*, *200* and occasionally *220* in the diffraction patterns of BP powder under pressure. No evidence for any phase transformation to another crystalline structure of BP, or an amorphous phase has been found over the pressure range studied. With increasing pressure, the diffraction lines of BP slightly broaden (FWHM is ~10% higher at 45 GPa as compared to ambient pressure), which is indicative of very low pressure-induced strain and quasi-hydrostatic compression. The good reproducibility of the lattice expansion data in two independent runs and well-resolved ruby signals in the entire pressure range under study additionally ensures the quasi-hydrostatic conditions of the EOS measurements.

Figure 1 shows the experimental $p$-$V$ equation of state of BP at 300 K for two powder and one single-crystal samples. A least-squares fit using the Murnaghan equation of state[27] gives the value of the bulk modulus $B_0 = 174(2)$ GPa with its pressure derivative $B_0' = 3.2(2)$. The zero-pressure unit cell volume was taken as $V_0 = 93.45$ Å$^3$. Fig. 1 also shows the theoretical equation of state that may be described by the Murnaghan equation with parameters $V_0 = 93.95$ Å$^3$, $B_0 = 165$ GPa, $B_0' = 3.3$, which are in good agreement with our experiments. The obtained value of the bulk modulus follows well Cohen's relationship between the bulk moduli and interatomic A–B distances of diamond-like $A^X B^{8-X}$ compounds ($B_0 = 165.3$ GPa);[28] and is in excellent agreement with the values obtained by elastic measurements ($B_0 = 173$ GPa)[29] and semiempirical estimations ($B_0 = \sim 180$ GPa).[30,31] The previously reported value of 267 GPa from non-hydrostatic experiment without pressure medium[32] seems to be highly overestimated.

## B. First and second order Raman spectrum of BP

At ambient conditions the Raman spectrum of BP (Fig. 2) exhibits two bands: a very weak at 794 cm$^{-1}$ and a strong one at 823 cm$^{-1}$ that correspond to the Brillouin zone center transverse TO and longitudinal LO optical modes, respectively. According to group theory, a zinc-blend lattice (*F-43m* space group (No. 216), with two independent atomic positions with Wyckoff notations 4a and 4b) give a rise to two zone-center Raman-active phonon modes, both of $T_2$ symmetry.[33]

Second-order Raman spectrum has been observed in the 1300-1700 cm$^{-1}$ frequency range (Insert of Fig. 2). According to group theory, the bands can combine the first-order frequencies from all $\Gamma$, K, L, Q and X points of Brillouin zone.[34,35] Experimental values of all these frequencies are not available so far, and their estimations we made by analogy with cBN, taking into account only the overtones. In the harmonic approximation, the stretching wavenumber is the square root of the ratio between bond stiffness, $k_{XY}$, and reduced mass, $M_{XY}$, i.e. $\nu_1 = (k_{1XY}/M_{XY})^{1/2}$ for TO and $\nu_2 = (k_{2XY}/M_{XY})^{1/2}$ for LO. Suggesting that the stiffness constant is the same for all Brillouin zone points of a given mode, one can easily obtaine the estimate for BP overtone optical modes by keeping $\nu_{1,cBN}/\nu_{1,BP} = \nu_{0,cBN}/\nu_{0,BP}$ (Tab. 1).

Our more detailed analysis (Fig. 3) were performed based on the available data on the lattice dynamics by *ab initio* calculations.[36] The empty up triangles in Fig. 3a show the TO, LO, TA and LA frequencies at different points of Brillouin zone, while solid up triangles in Fig. 3 - various combinations of these modes. The 1st order frequencies are LO($\Gamma$) at 825.0 cm$^{-1}$, TO($\Gamma$) at 800.0 cm$^{-1}$, LO(X) at 752.2 cm$^{-1}$, LA(X) at 480.0 cm$^{-1}$, TO(X) at 667.8 cm$^{-1}$, TA(X) at 304.8 cm$^{-1}$, LO(L) at 718.6 cm$^{-1}$, LA(L) at 480.4 cm$^{-1}$, TO(L) at 740.0 cm$^{-1}$, and TA(L) at 243.8 cm$^{-1}$.[36]

The vertical dashed lines indicate the most probable attribution of the observed second-order bands. Future studies using polarized Raman spectra are required to divide the second order into different symmetries, corresponding to either overtones or additive-subtractive combinations of first order modes from different high-symmetry critical points of the Brillouin zone and the phonon dispersion relations. This study, however, is out of the scope of the present work.

## C. Raman modes of BP under pressure.

Figure 4a shows the pressure evolution of the first-order Raman modes of BP. Non-linear pressure dependence of the Raman shift of the active modes has been observed. Raman frequencies were reproducible in three independent runs, as well as upon decreasing the pressure. The relation

$$\omega_p = \omega_0 \cdot \left(1 + \frac{\delta_0}{\delta'} \cdot p\right)^{\delta'} \qquad (1)$$

has been used to describe the pressure dependence of the modes. Derivation of this equation assumes that the logarithmic pressure derivative $\delta$ of the mode frequency varies linearly with pressure, $\delta = \delta_0 + \delta' \cdot P$, where $\delta_0 = (d\ln\omega / dP)_{P=0}$ and $\delta' = \delta_0^2 (d^2 \ln\omega / dP^2)^{-1}{}_{P=0}$ is the inverse value of pressure derivative of corresponding mode-hardening modulus $\delta_0^{-1}$, $\delta'$ being independent on pressure. A least-squares fit of relationship (1) to the experimental data yields the first- and second-order parameters $\delta_0 = 7.25(19) \cdot 10^{-3}$ GPa$^{-1}$ and $\delta' = 0.310(15)$ for TO mode ($\sim$797 cm$^{-1}$); and $\delta_0 = 6.50(23) \cdot 10^{-3}$ GPa$^{-1}$ and $\delta' = 0.265(36)$ for LO mode ($\sim$827 cm$^{-1}$). The corresponding absolute frequency shift is $\delta_0\omega_0 = 5.77$ and $5.38$ cm$^{-1}$/GPa for TO and LO modes respectively, that are higher than those obtained during the measurements in the methanol-ethanol pressure medium, i.e. 5.48 (TO) and 4.89 cm$^{-1}$/GPa (LO).[37] The mode Grüneisen parameters $\gamma_G = \delta_0 B_0$ (where $B_0 = 174$ GPa is the bulk modulus of BP) are $\gamma_G^{TO} = 1.26$ (e.g. $\gamma_G^{TO} = 1.257$[38] for cBN) and $\gamma_G^{LO} = 1.13$. Our results are consistent with the data obtained for other III-V diamond-like phases, for which $\gamma_G^{TO} > \gamma_G^{LO} \cong 1$.[37]

The pressure dependence of the ratio between the integral intensities of TO and LO Raman lines is shown in Fig. 4b. At low pressure the TO band is four times weaker as compared to LO, but at high enough pressures it becomes $\sim$2.5 times stronger ($I_{TO}/I_{LO}$ changes from $\sim$0.25 to $\sim$2.5). This observation is well reproducible in three different runs (Fig. 4b), on compression as well on decompression. Previously opposite situation (LO band is less intensive at low pressures) were reported for the first-order Raman scattering by TO and LO phonons in zinc blend GaSb, InAs, and InSb. The strong change of $I_{TO}/I_{LO}$ (e.g. from $\sim$3 down to $\sim$0.5 in the case of InAs) has been attributed to the resonance Raman scattering, where at some pressure the bandgap becomes the same as the energy of excitation beam.[14] However, this situation is not our case: the excitations beam was of 2.41 eV, which is close to the indirect bandgap at ambient pressure,[36] but quite different from the direct one of $\sim$3.5.[13] Even with pressure this bandgap cannot get a value close to indirect one.[12]

Previously the pressure dependence of the parameters that define the Raman scattering processes in compounds with zinc-blende structure has been studied.[15] According to these results, the difference in the intensities of LO and TO modes arises from the effective electro-optical constant $\alpha$ that defines LO mode and not TO. Under pressure, only $\alpha$ changes quite significantly, and so does $I_{TO}/I_{LO}$. According to the theoretical consideration,[15] $I_{TO}/I_{LO}$ and $\alpha$ follow the pressure dependence of molar volume $V$, the gap between relevant conduction band states ($\Gamma_{15c}$ and $\Gamma_{1c}$), and the difference in squared LO and TO frequencies:

$$\frac{I_{TO}}{I_{LO}} \sim \alpha^{-2} \sim V \cdot \frac{(E_{c15} - E_{c1})^2}{\omega_{LO}^2 - \omega_{TO}^2} \qquad (2)$$

Molar volume dependence is defined by the equation of state discussed above (Fig. 1) and the difference in squared LO and TO frequencies may be easily calculated using pressure dependencies of Raman shift (Fig. 4a). The gap between $\Gamma_{15c}$ and $\Gamma_{1c}$ lowest conduction states has been evaluated by our DFT calculations and may be described as $E_{c15}$ - $E_{c1}$ = 0.3191 + 6.24·$10^{-3}$ $p$ - 9.6·$10^{-6}$ $p^2$.

Figure 4b represents the impact of all three factors ($V$, [$E_{c15}$ - $E_{c1}$] and [$\omega_{LO}^2$ - $\omega_{TO}^2$]) scaled to match the zero-pressure intensity ratio. One can easily see that the evolution of $I_{TO}/I_{LO}$ can be satisfactorily explained by pressure dependence of $\alpha^2$. Both the gap between $\Gamma_{15c}$ and $\Gamma_{15c}$ conduction states and difference in squared LO and TO frequencies strongly decrease the $I_{LO}$ value as compared with $I_{TO}$.

## V. CONCLUSIONS

Using X-ray diffraction and Raman scattering, we have measured the changes in lattice parameters and Raman active modes of boron phosphide BP (TO and LO) up to 45 GPa. A least-squares fit to the volume-pressure data yields the bulk modulus $B_0$ = 174(2) GPa with the pressure derivative $B_0'$ = 3.2(2). The mode Grüneisen parameters, $\gamma_G^{TO}$ = 1.26 and $\gamma_G^{LO}$ = 1.13, are consistent with the corresponding data for other III-V diamond-like semiconductors, for which $\gamma_G^{TO} > \gamma_G^{LO} \cong 1$. Our results also indicate that the gap between $\Gamma_{15c}$ and $\Gamma_{1c}$ conduction states and difference in squared LO and TO frequencies, which strongly change under pressure, are responsible for the decrease the $I_{LO}$ value as compared to $I_{TO}$.

## ACKNOWLEDGMENTS

We thank Dr. Mukhanov for synthesis of BP powders, Prof. Edgar for providing BP single crystals, and Dr. Mezouar for assistance in high-pressure measurements at ESRF. Single-crystal X-ray diffraction study was carried out at beamline ID27 during beam time kindly provided by ESRF. High-pressure Raman experiments were performed at the Bayerisches Geoinstitut under the EU "Research Infrastructures: Transnational Access" Program (Contract No 505320 (RITA) – High Pressure). This work was financially supported by the Agence Nationale de la Recherche (grant ANR-2011-BS08-018) and DARPA (grant W31P4Q1210008). Calculations were performed on XSEDE facilities and on the cluster of the Center for Functional Nanomaterials, Brookhaven National Laboratory, which is supported by the DOE-BES under contract no. DE-AC02-98CH10086.

**Tab. 1**      Experimentally observed Raman modes of boron phosphide.

| Band | $\nu_{BP}$, cm$^{-1}$ experimental (theoretical) | $\nu_{cBN}$, cm$^{-1}$ Ref. [34] | $\nu_{cBN}/\nu_{BP}$ | $\nu_{BP}^{TO}$, cm$^{-1}$ prediction[34] | $\nu_{BP}^{LO}$, cm$^{-1}$ prediction[34] |
|---|---|---|---|---|---|
| TO ($\Gamma$) | ***794*** | 1055 | ***1.3287*** | - | - |
| LO ($\Gamma$) | ***823*** | 1304 | ***1.5845*** | - | - |
| 2TO(X) | 1360 | 1800 | 1.3287 | 1354.7 | - |
| 2LO(K) | - | 2170 | 1.5845 | - | 1369.5 |
| 2TO(K) | 1392 | 1830 | 1.3287 | 1377.3 | - |
| 2TO(Q) | | 1880 | 1.3287 | 1414.9 | - |
| 2LO(X, L) | 1434 | 2270 | 1.5845 | - | 1432.6 |
| 2TO(W) | 1453 | 1940 | 1.3287 | 1460.1 | - |
| 2TO(Q) | - | 2000 | 1.3287 | 1505.2 | - |
| 2TO($\Gamma$) | 1594 | 2110 | 1.3287 | 1588.0 | - |
| 2LO($\Gamma$) | 1643 | 2610 | 1.5845 | - | 1647.2 |

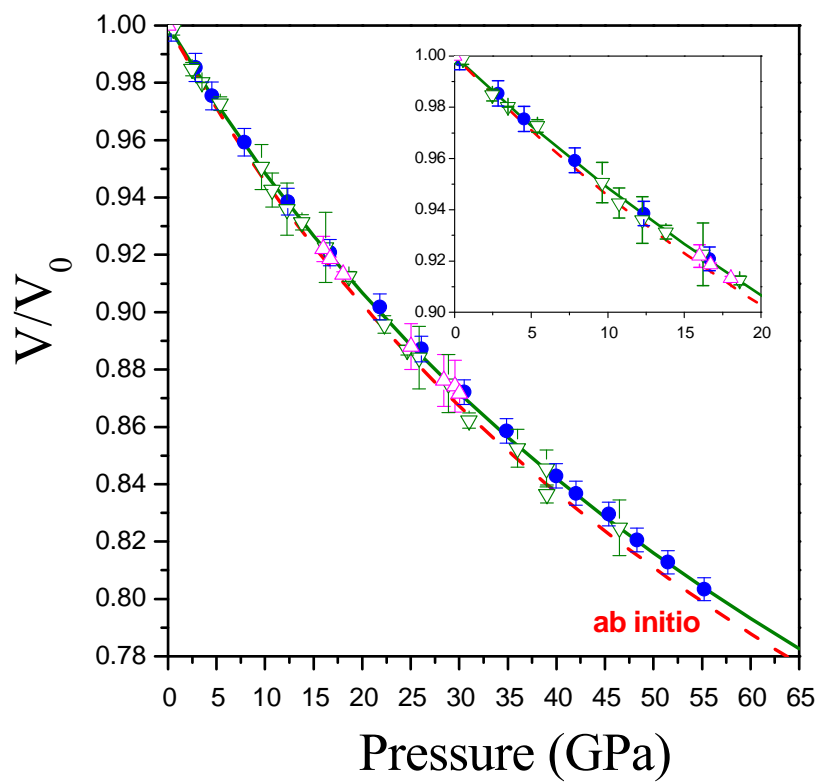

**Fig. 1** Equation of state of boron phosphide. Up and down triangles correspond to the powder X-ray diffraction data, while circles – to the single-crystal data. Dash line shows the results of *ab initio* calculations, solid line is the fit to the Murnaghan equation of state.

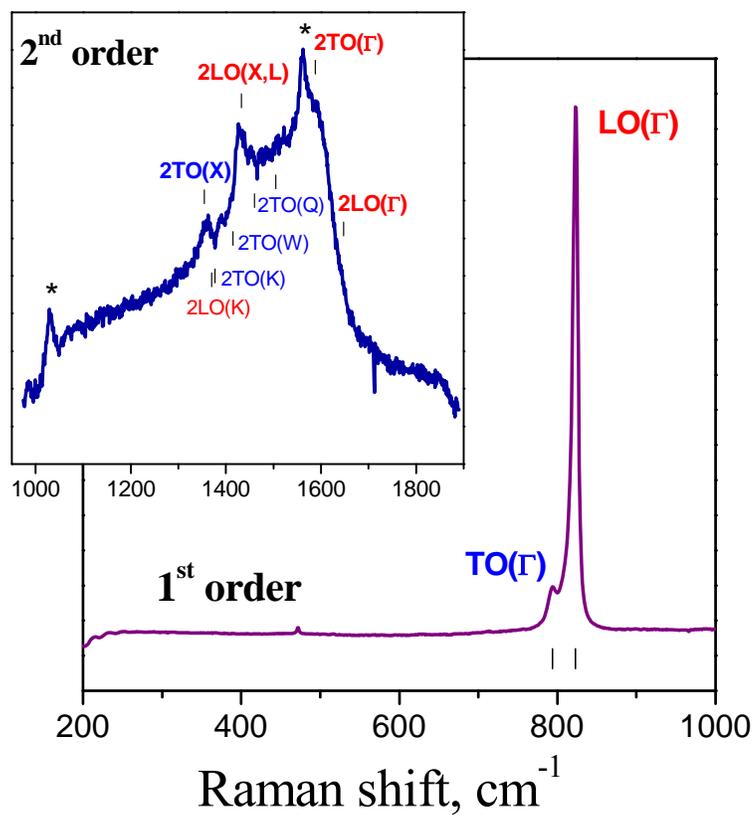

**Fig. 2** 1st and 2nd order Raman spectra of boron phosphide. The vertical bars show the overtone band positions.

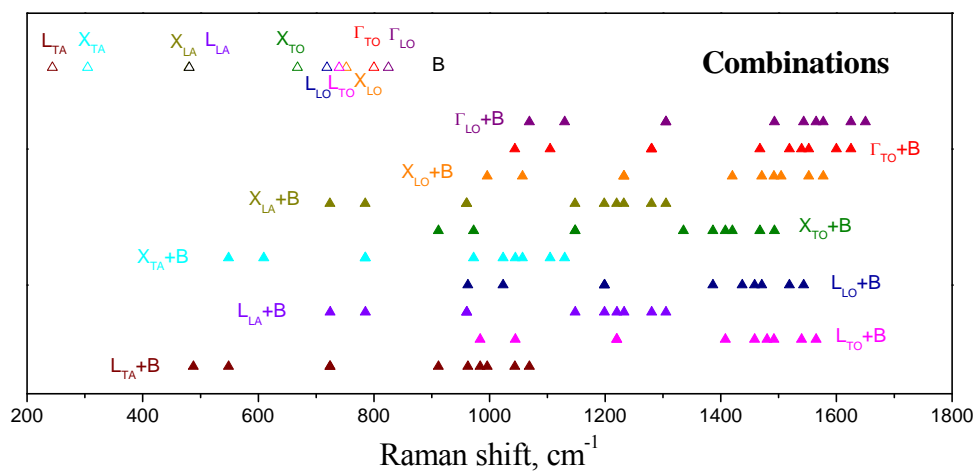

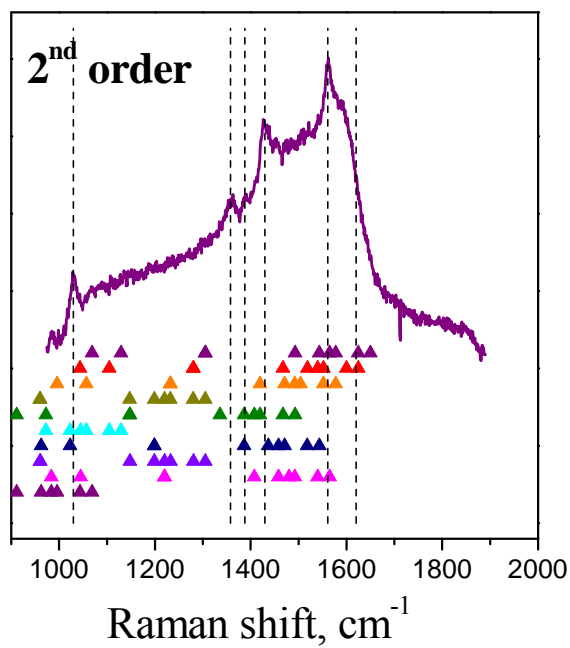

**Fig. 3** Second order Raman scattering of boron phosphide. The empty up triangles shows the TO, LO, TA and LA frequencies (all bands, B) at different points of Brillouin zone, while solid up triangles - various combinations of these modes **(a)**. The bands B were taken from *ab initio* calculations[36]: TA(L) - 243.8 cm$^{-1}$, TA(X) - 304.8 cm$^{-1}$, LA(X) at 480.0 cm$^{-1}$, LA(L) at 480.4 cm$^{-1}$, TO(X) at 667.8 cm$^{-1}$, LO(L) at 718.6 cm$^{-1}$, TO(L) at 740.0 cm$^{-1}$, LO(X) at 752.2 cm$^{-1}$, TO($\Gamma$) - 800.0 cm$^{-1}$, and LO($\Gamma$) - 825.0 cm$^{-1}$. The dashed lines indicate the most probable attribution of the observed second-order bands **(b)**.

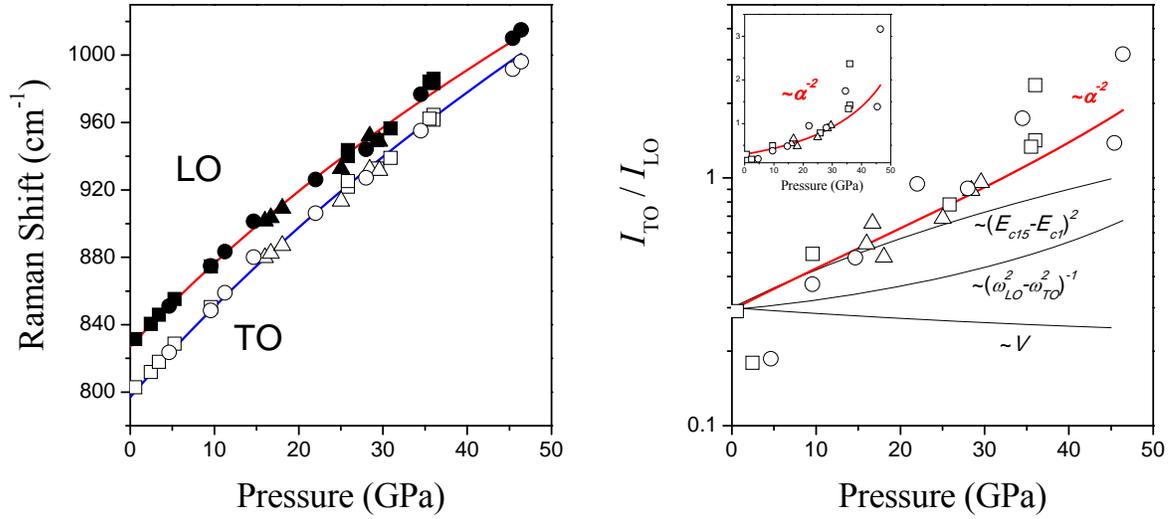

**Fig. 4** **(a)** Pressure dependence of the Raman shift of the observed TO and LO modes of BP. Circles, squares and triangles represent the results of three different runs. Solid lines represent the fit to eq. (1). **(b)** Pressure dependence of the $I_{\text{TO}}/I_{\text{LO}}$ ratio between the Raman intensities of the TO and LO modes of BP. Circles, squares and triangles represent the results of different runs. Thin solid curves show the high-pressure behavior of molar volume $V$, the gap between relevant conduction band states ($\Gamma_{15c}$ and $\Gamma_{1c}$), and the difference in squared LO and TO frequencies. Thick curve corresponds to the scaled inverse square of the effective electro-optical constant $\alpha$ that make different the HP behavior of LO and TO modes.